\documentclass[aps,prd,twocolumn,groupedaddress,showpacs,amsmath,amssymb]{revtex4-1}

\usepackage{graphicx}
\usepackage{bm,color}

\newcommand{\be}{\begin{eqnarray}}
\newcommand{\ee}{\end{eqnarray}}

\newcommand{\hth}{\hat{\theta}}

\def\hb{\hat{b}}
\def\hn{\hat{\rho}}

\begin{document}

\title{Glassy Dynamics from Quark Confinement: \\
Atomic Quantum Simulation of Gauge-Higgs Model on Lattice
}
\author{Jonghoon Park$^1$}
\author{Yoshihito Kuno$^2$}
\author{Ikuo Ichinose$^1$}
\affiliation{$^1$Department of Applied Physics, Nagoya Institute of Technology, Nagoya, 466-8555, Japan}
\affiliation{$^2$Department of Physics, Graduate School of Science, Kyoto University, 
Kyoto, 606-8502, Japan}

\date{\today}

\begin{abstract}
In the previous works, we proposed atomic quantum simulations of the U(1)
gauge-Higgs model by ultra-cold Bose gases.
By studying extended Bose-Hubbard models (EBHMs) including long-range repulsions,
we clarified the locations of the confinement, Coulomb and Higgs phases.
In this paper, we study the EBHM with nearest-neighbor repulsions 
in one and two dimensions at large fillings by the Gutzwiller variational method.
We obtain phase diagrams and investigate dynamical behavior of electric flux
from the gauge-theoretical point of view. 
We also study if the system exhibits glassy quantum dynamics in the absence and 
presence of quenched disorder.
We explain that 
the obtained results have a natural interpretation in the gauge theory framework.
Our results suggest important perspective on many-body localization in  
strongly-correlated systems.
They are also closely  related to anomalously slow dynamics observed by 
recent experiments performed on Rydberg atom chain, and our study indicates existence
of similar phenomenon in two-dimensional space.
\end{abstract}

\maketitle


\section{Introduction}{\label{intro}}

Ultra-cold atomic gas systems are one of the most actively studied subjects
in physics these days~\cite{Georgescu}.
By their high-controllability and versatility, the ultra-cold atoms provide an important
playground for study on interesting problems in quantum physics.
In particular, dynamical properties of the many-body quantum systems can be
investigated by controlling physical parameters of the systems. 
Most of these investigations are beyond the reach of the conventional research methods
such as various numerical methods including the Monte-Carlo simulations, 
density-matrix renormalization group, etc.
From this point of view, the ultra-cold atom systems are sometimes called
ideal quantum simulators~\cite{book,Blochrev}.

Among them, numerous interesting studies on quantum simulations of
the lattice gauge theory (LGT) have been reported
\cite{Zohar1,Zohar2,Tagliacozzo1,Banerjee1,Zohar3,Zohar4,Banerjee2,Tagliacozzo2,Zohar5,Wiese,Zoharrev,Bazavov,GHcirac,Zn,Schwinger,string}.
Various setups using internal degrees of freedoms of atoms have been proposed.
In these studies, one of the most important point is how to realize the local
gauge symmetry in charge-neutral atomic systems.
In the previous works~\cite{ours1,ours2,ours3,ours4}, we considered single-component 
Bose gas systems described by an extended Bose-Hubbard model (EBHM) \cite{Dutta},
and show that the U(1) gauge-Higgs model  with the exact local gauge symmetry
can be quantum simulated by the EBHM.
The gauge-Higgs model (GHM)  is one of the most fundamental gauge theories \cite{Fradkin,Kogut} 
in not only high-energy physics but also condensed matter physics.
The GHM has (at least) two distinct phases, one is the confinement
phase and the other is the Higgs phase.
In our works, we clarified phase diagrams by using the Monte-Carlo (MC)
simulations.
Dynamical variables such as the electric field exhibit very different behaviors
in the above two phases, and we studied their dynamics by using 
the Gross-Pitaevskii equations.

In this paper, we continue the above study and investigate the 
EBHM and GHM by the Gutzwiller (GW) variational method.
In particular, we are interested in case of relatively large fillings with the average
particle number per site $\rho_0=7\sim 30$, as
large filling legitimates the use of the GW variational method and the EBHM-GHM
correspondence.

This paper is organized as follows.
In Sec.~II, we introduce the EBHM and explain how it quantum simulates
the GHM on the lattice.
We also briefly summarize the previous works.
In Sec.~III, we show the numerical results for the model in one and two dimensions.
We first clarify the phase diagrams of the EBHM,
and identify the parameter regions corresponding to the confinement and Higgs phases.
Then, we investigate the dynamical behavior of the electric flux put in the central
region of the lattice.
In the confinement phase, the electric flux is stable although it exhibits 
string-breaking-like fluctuations.
On the other hand in the Higgs phase,
it spreads in the empty space and breaks into bits.
This result is in good agreement with the previous result obtained by 
the Gross-Pitaevskii equations.
In Sec.~IV, we study the robustness of confinement state in the GHM and 
the effect of the random chemical potential on it.
In particular, we observe a kind of glassy dynamics of configurations
with a finite synthetic electric field in the confinement phase.
This behavior is reminiscent of anomalously slow dynamics observed by 
recent experiments performed on Rydberg atom chain~\cite{Rydberg}
as indicated by Ref.~\cite{string}.
Then, it is interesting to study the effect of quenched disorder induced by 
the random chemical potential on the glassy state.
We calculate life time of high-energy states with density-wave (DW)-type configurations
for various the strength of the disorder.
We obtain somewhat `unexpected' results, that it, a weak disorder hinders 
the glassy state first, whereas further increase of disorder enhances the glassy nature.
This means that there exists a critical strength of the disorder at which 
the glassy nature is hindered maximally.
Section V is devoted for discussion and conclusion.
We discuss the observed glassy behavior of the confinement phase from 
the gauge-theoretical point of view, and clarify the origin of the above `unexpected'
results.
We also suggest certain experiments for examining our observation and 
searching many-body localization (MBL) in ultra-cold gases with a dipole moment.

\section{Extended Bose-Hubbard model and Gauge-Higgs Model}

In the previous works~\cite{ours1,ours2,ours3,ours4}, 
we showed that the GHM appears as a low-energy
effective theory from the EBHM.
For the simplicity of the presentation, here we consider the one-dimensional (1D) EBHM,
and explain its relation to the GHM.
Extension to higher-dimensional cases are rather straightforward although 
long-range repulsions are necessary.
Hamiltonian of the EBHM in 1D is given as follows,
\be
H_{\rm EBH}&=&-J\sum_i(\hb^\dagger_i\hb_{i+1}+\hb^\dagger_{i+1}\hb_i)
+{U \over 2}\sum_i\hn_i(\hn_i-1) \nonumber \\
&&+V\sum_i\hn_i\hn_{i+1}-\mu\sum_i\hn_{i},
\label{HEBH}
\ee
where $\hb_i \ (\hb^\dagger_i)$ is the boson annihilation (creation) operator at site $i$,
$\hn_i=\hb^\dagger_i\hb_i$, and $\mu$ is the chemical potential.
The $U$-term and $V$-term in Eq.~(\ref{HEBH}) are one-site and nearest-neighbor
(NN) repulsions, respectively.
We introduce the phase ($\hth_i$) operator as follows, 
$\hb_i=e^{i\hth_i}\sqrt{\hat{\rho}_i}$.
By controlling the chemical potential, we consider the case of relatively large fillings 
such as $\rho_0={1 \over L}\sum_i\langle \hat{\rho}_i\rangle=(7\sim 30)$ in this paper,
where $L$ is the linear system size. 
To relate the boson operator to the gauge field, we introduce a dual lattice with 
site $r$, which corresponds to link $(i,i+1)$ of the original lattice.
Artificial electric field, $E_r$, and vector potential, $A_{r,1}$, are given by
$E_r=-(-)^r(\hat{\rho}_i-\rho_0)\equiv -(-)^r\eta_r$, and $A_{r,1}=(-)^r\hth_i$.
It is verified that $E_r$ and $A_{r,1}$ satisfy the ordinary canonical commutation relations 
such as $[E_r, A_{r',1}]=-i\delta_{rr'}$.
Then, the following Hamiltonian is derived from $H_{\rm EBH}$ [Eq.~(\ref{HEBH})]
by ignoring the third or higher-order terms of $\{\eta_r\}$ as we do not
consider the system in the critical regimes,
\be
H_{\rm GH}&=&\sum_r\Big[{V\over 2}(E_{r+1}-E_r)^2+{g^2 \over 2}E^2_r
\nonumber \\
&&-2J\rho_0\cos (A_{r+1,1}+A_{r,1})\Big],
\label{HGH}
\ee
where $g^2=U-2V$.

From Eq.~(\ref{HGH}), the partition function of the system, $Z$, is given by
the imaginary-time path integral,
\be 
Z&=&\int [dA_1][dE] \nonumber \\
&&\times \exp\Big[\sum_\tau(iE_x(A_{x+0,1}-A_{x,1})
-\Delta\tau H_{\rm GH})\Big],
\label{ZGH}
\ee
where we have introduced the imaginary time $\tau$ and the corresponding
lattice with time slice $\Delta\tau$.
Now, the system in Eq.~(\ref{ZGH}) is defined on 2D lattice with site 
$x=(x_0,x_1)=(\tau,r)$, and $x+0=(\tau+1,r), \ x+1=(\tau,r+1)$.
It is obvious that the Hamiltonian $H_{\rm GH}$ in Eq.~(\ref{HGH}) 
and the partition function $Z$ in Eq.~(\ref{ZGH}) are {\em not} invariant
under a local gauge transformation such as 
$A_{x,1}\to A_{x,1}-\nabla_1\alpha_i$,
where $\nabla_1\alpha_i=\alpha_{i+1}-\alpha_i$ $[r=(i+1,i)]$ and $\{\alpha_i\}$ are arbitrary
real parameters at original sites.
In Ref.~\cite{ours1}, we showed that the system given by Eqs.~(\ref{HGH}) and (\ref{ZGH})
can be regarded as the U(1) GHM with the {\em exact local gauge symmetry}.
In order to express the partition function $Z$ in a gauge-invariant form, 
we introduce two-component compact gauge potential on the link $(x,x+\nu) \ (\nu=0,1)$,
$U_{x,\nu}=e^{iA_{x,\nu}}$ and Higgs field $\phi_x=e^{i\varphi_x}$.
Then, we can prove the following equation,
\be
&&Z=\int[dA_0][dA_1][d\phi]\exp [A_{\rm GH}],  \nonumber \\
&&A_{\rm GH}=A_I+A_P+A_H, \nonumber \\
&&A_I={1 \over 2V\Delta\tau}\sum_x\bar{\phi}_{x+0}U_{x,0}\phi_x+\mbox{c.c.}, 
\label{ZGH2} \\
&&A_P={1\over 2g^2\Delta\tau}\sum_x \bar{U}_{x,0}\bar{U}_{x+0,1}
U_{x+1,0}U_{x,1}+\mbox{c.c.}, \nonumber \\
&&A_H=J\rho_0\Delta\tau\sum_x\bar{\phi}_{x+2}U_{x+1,1}U_{x,1}\phi_x
+\mbox{c.c.},
\nonumber
\ee
where $A_I$ is the hopping term of the Higgs field in the $\tau$-direction
(the kinetic term), $A_P$ is the plaquette term of the gauge field 
(the electro-magnetic term), and $A_H$ is the spatial hopping term of the Higgs field.
The time-component of the gauge field $A_{x,0}$ has been introduced as an
auxiliary field in order to perform the integration over the electric field $E_r$.
It is easily to show that the system described by Eq.~(\ref{ZGH2}) is gauge-invariant.
By fixing the gauge freedom with the gauge condition such as $\phi_x=1$,
which is so-called unitary gauge, the system Eq.~(\ref{ZGH2}) reduces to the one
derived from the original system Eq.~(\ref{ZGH}) by integrating out $E_r$ with 
the auxiliary field $A_{x,0}$. 
From the action in Eq.~(\ref{ZGH2}), it is shown that
for large $J\rho_0$, the Higgs phase is realized, whereas the (homogeneous)
confinement phase forms for large $U, V$ and $g^2>0$.

In the previous work~\cite{ours4}, 
we investigated the phase diagrams of the EBHM 
[Eq.~(\ref{HEBH})] and the GHM [Eq.~(\ref{ZGH2})] by means of the MC simulations
separately, and verified that the phase diagrams of two models are consistent
with each other.
There exist three phases in the phase diagram, i.e., the superfluid (SF), 
Mott insulator (MI) and DW.
It was shown that the SF corresponds to Higgs phase of the gauge theory,
whereas the MI in the vicinity of the DW corresponds to the confinement phase 
of the gauge theory.
We also studied the 2D and 3D EBHM from the view point of a quantum simulation for
the lattice GHM, and obtained interesting results~\cite{ours1,ours2,ours3}.
In this paper,
we shall study the EBHM in 1D and 2D at relatively large fillings by means of the GW variational method.
At large fillings, the GW variational method is reliable even for the 1D system, as 
it is expected that a quasi-Bose-Einstein condensation forms {\em at each site} 
of the optical lattice at large fillings and the GW variational method can describe
dynamics of both the MI and SF.

\begin{figure}[t]
\centering
\begin{center}
\includegraphics[width=7cm]{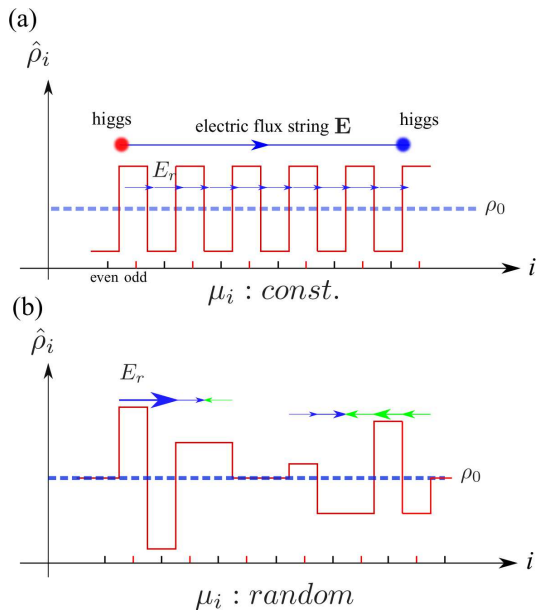}
\end{center}
\caption{(a) Stable electric flux string: the blue dashed line represents background 
mean density. 
The mean density is constant.
The blue arrows and the red line represent the electric fields and $\rho_{i}$, respectively.
The electric flux string corresponds to the DW pattern of the density modulation. 
At edges of the flux string, the fictitious Higgs charges appear. 
In the confinement phase in 2D and 3D systems, electric flux forms a straight line.
(b) Random chemical potential case:
The random chemical potential leads to a random mean density pattern. 
The short electric-flux string are generated.  
The blue and green arrows represent the short flux strings.
Existence of random short electric fluxes induces an instability of the long electric 
flux string shown in the upper panel (a). 
Here, the Higgs charges residing at the edges of the short electric string are omitted. 
}
\label{electricflux}
\end{figure}

Before going into the numerical study, it is useful to review the relation of the EBHM and 
GHM in a pictorial way.
As explained above, Mott state in the vicinity of the DW corresponds to the confinement
phase.
In the atomic perspective, one-dimensional DW-type configuration of atoms 
with a density modulation is an interesting object.
It is expected that such a configuration has a rather long lifetime because of 
the NN interactions as observed by recent experiment on Rydberg atom~\cite{Rydberg}.
This configuration is nothing but an electric flux string in the gauge-theory perspective.
As schematically shown in Fig.~\ref{electricflux} (a), a pair of Higgs particle are attached
to edges of the electric flux as dictated by the Gauss law.
Even in 2D system, the electric flux string tends to form a straight line in the confinement
phase, and it can change its length only by pair creation of Higgs particle. 
Let us consider effects of background density fluctuations around the DW string,
which are generated by a random chemical potential, see Fig.~\ref{electricflux} (b). 
In the gauge-theory picture, these fluctuations correspond to randomly distributed
charges and resultant electric-field fluxes, which induce an instability of the original
electric flux string.
Through this picture, we expect that the lifetime of the DW string is shorten by 
the random chemical potential.
This may be an unexpected result from the common brief that a high-energy 
state such as a DW string is stabilized by disorders as a result of localization, 
but is quite natural from the gauge-theoretical point of view.
In the subsequent section, we shall verify the above gauge-theoretical expectation
by the numerical simulations.


\section{Numerical Results: Systems without disorder}

In this section, we show the numerical results for the EBHM in 1D and 2D 
obtained by the GW variational method.
The Hamiltonian of the EBHM in Eq.~(\ref{HEBH}) is factorized into 
single-site local Hamiltonian with the maximum particle number at each site,
$n_c$~\cite{FN1}.
In this work, we set $n_c=30$ for 1D and $n_c=50$ for 2D systems. 
While so far the 1D EBHM has been extensively studied under unit filling condition
\cite{Rossini,Batrouni,Kawaki}, our focus is large filling regime, thus it is worth
characterizing the large filling ground state.
We also employ the periodic boundary condition for the practical calculation.

We first study the phase diagrams and identify the parameter regions of
the confinement and Higgs phases.
Then, we investigate dynamical properties of the gauge field in these phases.

\subsection{Phase diagram of 1D EBHM}

In this subsection, we study equilibrium properties of the system,
in particular, the ground-state phase diagrams of the 1D EBHM.
To this end, we obtain the lowest-energy states for $H_{\rm EBH}$
by the GW variational method. 
Order parameters, which are used for identification of phases, the following;
\be
\Phi={1 \over N_s}\sum_i\Phi_i={1 \over N_s}\sum_i\langle \hb_i\rangle, \;\; 
\Delta n=\bar{\rho}_e-\bar{\rho}_o,
\label{OPs}
\ee
where $\bar{\rho}_{e(o)}$ is the average density of atom at even (odd) sites,
and $N_s$ is the total number of sites and $N_s=L=200$ in the present calculation.
Finite value of $\Phi$ indicates the existence of the SF, and $\Delta n$
measures the DW.
As we fix the chemical potential ($\mu$) in the calculation, the total average density 
of atom, $\rho_0$, varies under a change of the parameters in the Hamiltonian 
$H_{\rm EBH}$.

\begin{figure}[t]
\centering
\begin{center}
\includegraphics[width=8cm]{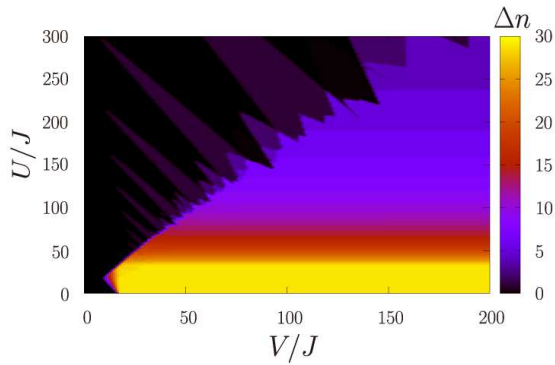}
\end{center}
\vspace{-1cm}
\caption{$\Delta n$ in the $(V/J-U/J)$ plain for the 1D EBHM.
There are four phases, Mott insulator (MI), superfluid (SF), density wave (DW),
and supersolid (SS).
Measurement of the SF order is shown in Fig.~\ref{SF1d}.
$\mu/J=950$.
}
\label{DW1d}
\end{figure}

\begin{figure}[t]
\centering
\begin{center}
\includegraphics[width=8cm]{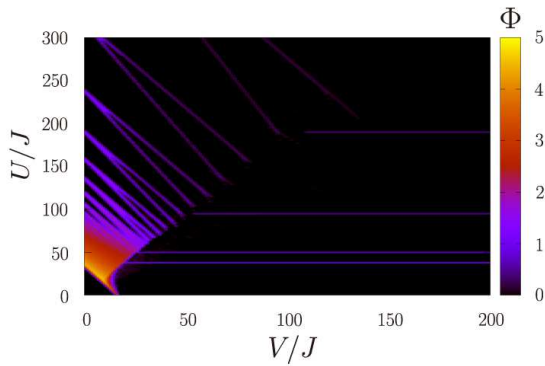}
\end{center}
\vspace{-1cm}
\caption{$\Phi$ in the $(V/J-U/J)$ plain for the 1D EBHM.
In the MI and DW, $\Phi=0$, whereas in SF and SS, $\Phi>0$.
For small $V/J, U/J$, the SF order parameter is vanishingly small.
This result comes from the fact that the particle number at each site
saturates the maximum value $n_c$, and the GW method is not applicable there. 
}
\label{SF1d}
\end{figure}

\begin{figure}[t]
\centering
\begin{center}
\includegraphics[width=8.5cm]{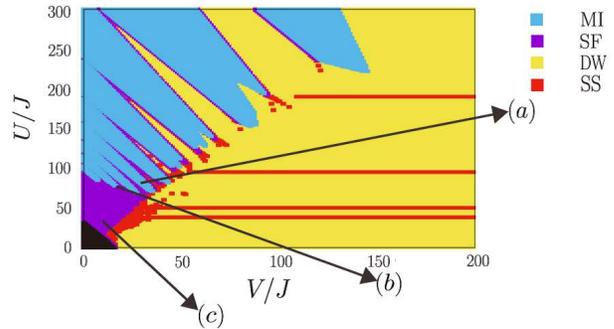}
\end{center}
\vspace{-0.5cm}
\caption{Phase diagram of the 1D EBHM.
For large $U/J, V/J$, the MI and DW occupy the phase diagram, and 
the SF is located between the MI and DW phases.
The arrows indicate the locations in which the behavior of electric flux is measured.
}
\label{1DPD}
\end{figure}

In Fig.~\ref{DW1d}, we show the calculation of $\Delta n$ in the $(V/J-U/J)$-plain.
$J=0.01$ and chemical potential is fixed as $\mu/J=950$ to obtain relatively large fillings.
In Fig.~\ref{SF1d}, we show the calculation of $\Phi$.
From the results in Figs.~\ref{DW1d} and \ref{SF1d}, we obtain the phase
diagram of the 1D EBHM as in Fig.~\ref{1DPD}~\cite{QMC}.
SF forms in the regions of relatively small $U/J$ and $V/J$.
MIs for large $U/J$ have large integer filling factors such as $\rho_0=7$ for 
$U/J=71$ and $V/J=35$.


In Fig.~\ref{1DPD}, the three parameter regions indicated by the arrows 
refer to the confinement [(a)], Higgs close to confinement [(b)], 
and genuine Higgs phases [(c)], respectively.
Here, we should comment that in Fig.~\ref{SF1d}, there are many lines where 
the finite SF density appears. 
These lines exist between the MIs with different fillings or the DW phase. 
Supersolid (SS) also exists in some parameter regions including narrow line regions
between the MIs and DW.
Similar tendency was reported in Ref.~\cite{Batrouni}.
In the subsequent section, we shall study physical properties of the above phases
from the viewpoint of the gauge theory.

\subsection{Behavior of electric flux in quantum simulation of gauge-Higgs model:
1D case}

In this subsection, we shall study the time evolution of 
``electric flux" put on a straight line.
To this end, we employ the time-dependent GW
methods~\cite{tGW1,tGW2,tGW3,tGW4,tGW5,tGW6,aoki,ours5}.
Behavior of the electric flux is a very important quantity in the gauge theory,
which discriminates the confinement, Coulomb and Higgs phases \cite{ours1}.
In the EBHM, an artificial electric flux at $r$ is produced by the configuration 
such as 
$\langle E_r\rangle=-(-)^r(\langle \hat{\rho}_i\rangle-\rho_0)=\Delta$, 
where $\Delta$ specifies
a pair of charge, $(-\Delta,+\Delta)$, located at the edges of the electric flux string.
In the GHM, this configuration is explicitly given by
$
\prod_{r_1<r<r_2}(U_{r,1})^\Delta|0\rangle,
$
where a pair of static charge $\pm \Delta$ are located at $r_1$ and $r_2$
and $|0\rangle$ is the `vacuum' without electric fluxes.
In the practical calculation, we add very small but finite fluctuations in local 
density of boson (i.e., local electric field) for initial states in order to perform smooth 
calculations by the time-dependent GW method.
In most of practical simulations in this section, we put $J=0.01$ and
set unit of time with $\hbar/(100J)$, i.e., we use unit of energy to set unit of time.

We consider the 1D case.
In the phase diagram shown in Fig.~\ref{1DPD}, we exhibit three typical parameter
regions corresponding to (a) MI in the vicinity of the DW ($U/J=71, \ V/J=35$), 
(b) SF close to MI ($U/J=70, \ V/J=14$), and (c) SF ($U/J=27, \ V/J=10$).
For all cases, $J=0.01$ and $\mu/J=950$.
System size $N_s=200$, and the electric flux is put from $r=70$ to $r=130$.
The confinement phase corresponds to the case (a), and the Higgs phase 
to the case (c).

For the case (a), we performed numerical simulations for two cases, i.e.,
the first one for the background particle density $\rho_0=10$ and 
the magnitude of the electric flux $\Delta=3$, and the second one for 
$\rho_0=7$ and $\Delta=1$.
The equilibrium filling of the MI at this parameter is $\rho_0=7$. 
As we explained above, this parameter region corresponds to the confinement
phase, and therefore we expect that the electric flux is rather stable and remains in 
the original position without breaking up small pieces for rather long period.

\begin{figure}[t]
\centering
\begin{center}
\includegraphics[width=7cm]{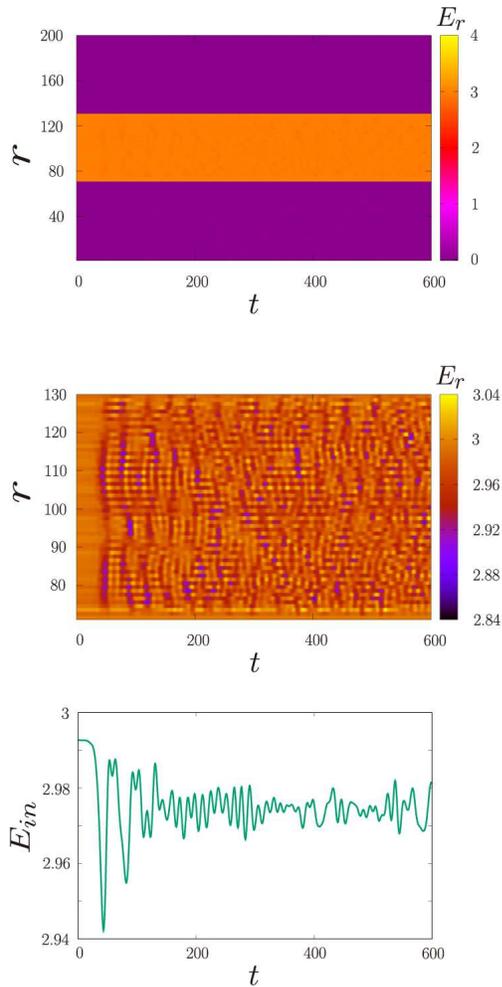}
\end{center}
\vspace{-0.5cm}
\caption{Upper panel: Time evolution of electric flux located in the center of 
the 1D system in the confinement phase.
$\rho_0=10$ and $\Delta=3$.
Electric flux is quite stable.
Middle and lower panels: Small but finite fluctuations of electric flux are observed.
In the lower panel, $E_{\rm in}(t=0)\neq 3$ comes from the density fluctuation
of the initial state that we employed.
}
\label{EF1d}
\end{figure}

In Fig.~\ref{EF1d}, we show the results of the simulation for $\rho_0=10$ and $\Delta=3$.
The electric flux is stable as we expected.
Close look at the inside of the electric flux reveals that small but finite fluctuations
of the electric field take place there~\cite{ours4}.
We studied the fluctuations of the electric field in the central region, 
\be
E_{\rm in}\equiv {1 \over N_i}\sum_{70\leq r \leq 130} E_r,
\label{Ein}
\ee
where $N_i$ is the length of the initial electric string,
and the result
is shown in the middle and the lower panels in Fig.~\ref{EF1d}.
Averaged electric field first decreases slightly, and then keeps constant with small fluctuations.
In the gauge theoretical point of view, the stability means that the system is in confinement phase as we expected.

Recently, closely related experiments were done on Rydberg atom 
chains~\cite{Rydberg}.
By the strong NN repulsion between Rydberg states, the system is nearly unit-filling,
and the DW type configurations exhibit anomalous slow dynamics.
In Ref.~\cite{string}, this phenomenon is interpreted as reminiscence of  
string-breaking of electric flux in the confined gauge theory~\cite{ours4,nonAbelian}. 
We will discuss this gauge-theoretical interpretations somewhat in detail in Sec.~V.

This stability of the electric field implies that the original EBHM exhibits a glassy 
behavior in the parameter
region corresponding to the confinement phase of the corresponding GHM.
This observation will be examined in the subsequent section.

We also performed numerical calculations for $\rho_0=7$ and $\Delta=1$.
The obtained results are quite similar to those for $\rho_0=10$ and $\Delta=3$
in Fig.~\ref{EF1d}.
The electric flux is quite stable even for $\Delta=1$, as the background particle density,
$\rho_0=7$, is equal to that of the equilibrium value.

\begin{figure}[t]
\centering
\begin{center}
\includegraphics[width=6cm]{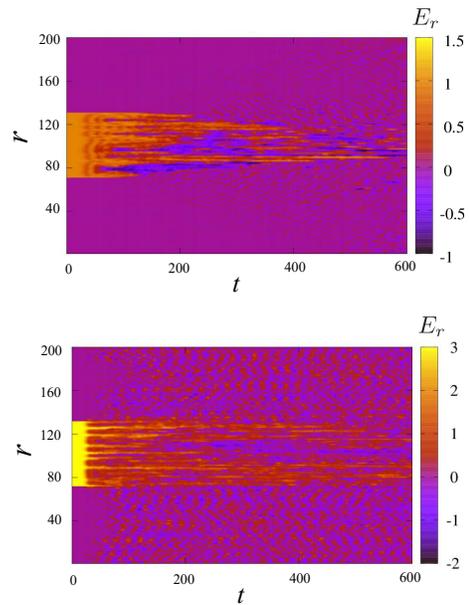}
\end{center}
\caption{
Upper panel: Time evolution of electric flux located in the center of the 1D system.
$\rho_0=10$ and $\Delta=1$.
The system exists in the SF close to the MI, which corresponds to the Higgs phase
relatively close to confinement.
Lower panel: Time evolution of electric flux located in the center of the 1D system
in the Higgs (SF) phase.
$\rho_0=20$ and $\Delta=3$.
}
\label{EF1dSF1}
\end{figure}

Let us turn to case (b).
We show the numerical results in Fig.~\ref{EF1dSF1}.
It is obvious that the electric flux string keeps the original configuration for a while, but it breaks into small pieces and these pieces spread the whole system.
This indicates the instability of the electric flux.  
In the Higgs phase of the gauge theory, electric charge is {\em not conserved}, and 
the electric fluxes are destroyed and also generated in various places.

Finally, we show the evolution of the electric flux in the case (c) in Fig.~\ref{EF1dSF1}.
It is obvious that the electric flux decays quite easily, and the whole system is full of large
fluctuations of electric field.
This means that the system is in deep Higgs phase.

\begin{figure}[t]
\centering
\begin{center}
\includegraphics[width=7.5cm]{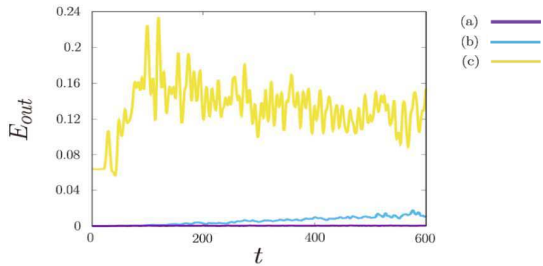}
\end{center}
\vspace{-0.5cm}
\caption{Time evolution of electric flux in the outside region.
Cases (a), (b) and (c). 
For the case (a), $E_{\rm out}(t=0)\neq 0$ comes from the density fluctuation of 
the initial state.
}
\label{Eout}
\end{figure}

In order to verify the above behavior of the electric flux, we measured average of 
electric field in the outside of the original location of the electric flux, i.e., 
\be
E_{\rm out}\equiv {1 \over N_o}\sum_{0<r<70, 130<r<200}E^2_r,
\label{defEout}
\ee
where $N_o$ is the number of sites in which the electric fluxes do not exist in the
initial configuration.
We show the results in Fig.~\ref{Eout}.
It is obvious that $E_{\rm out}$ is getting larger for smaller $V/J$ as we expected.

Let us briefly comment on the SS.
In the SS phase, phase coherence is remained,
but the density fluctuation in the SS is smaller than that in the SF regime. 
The density-wave configuration in the SS phase does not affect to 
the back-ground charge in the sense of gauge theory because the 
{\em density fluctuation itself} corresponds to electric fields. 
In this sense, the SS phase possesses Higgs-phase-like properties~\cite{ours4}.


\subsection{Phase diagram of 2D EBHM and behavior of electric flux}

\begin{figure}[t]
\centering
\begin{center}
\includegraphics[width=6.5cm]{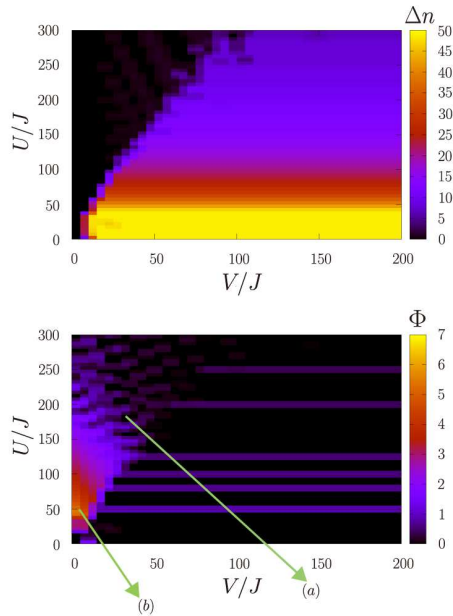}
\end{center}
\vspace{-0.5cm}
\caption{DW order (upper panel) and SF (lower panel) in the $(V/J-U/J)$ plain.
$J=0.01$ and $\mu/J=2000$.
(a) ((b)) corresponds to confinement (Higgs) phase.}
\label{2DPD}
\end{figure}

In this subsection, we shall study the 2D EBHM and 2D GHM.
We first show the phase diagram of the 2D EBHM at large fillings obtained by
the GW methods.    
Used order parameters are the superfluidity, $\Phi$, and DW, $\Delta n$ as in
the study of the 1D system.
The obtained numerical calculations and phase diagram are shown in Fig.~\ref{2DPD},
and we also indicate the parameter regions in which stability of the electric
flux will be examined.
As in the 1D case, the MI and DW occupy most of the phase diagram for
large $U/J$ and $V/J$,
and the SF forms in narrow regions between the MI and DW.
Most of the calculations were performed for the system size $N_s=20\times 20$.

\begin{figure}[h]
\centering
\begin{center}
\includegraphics[width=7cm]{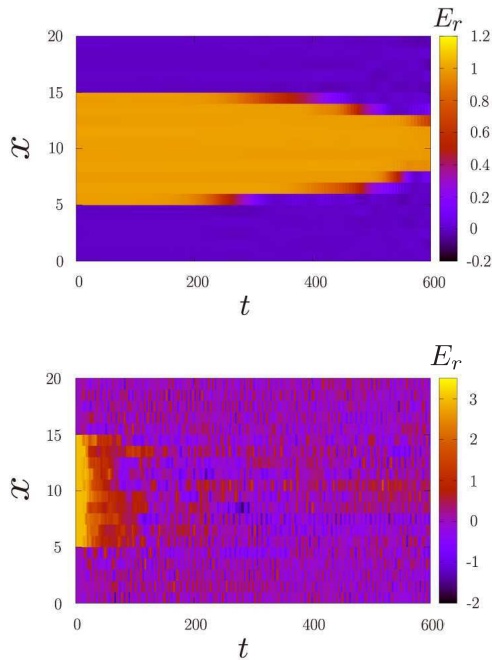} 
\end{center}
\vspace{-0.5cm}
\caption{Time evolution of electric flux put in the center of the system.
Upper panel: Confinement region with $\rho_0=7$ and $\Delta=1$.
Electric flux is stable for long period.
Lower panel: Higgs region with $\rho_0=30$ and $\Delta=3$.
Electric flux decays rapidly.
The on-site and nearest-neighbor repulsions play an essential role for
the stability of electric flux.
}
\label{ectric2DelF}
\end{figure}
\begin{figure}[h]
\centering
\begin{center}
\includegraphics[width=5.7cm]{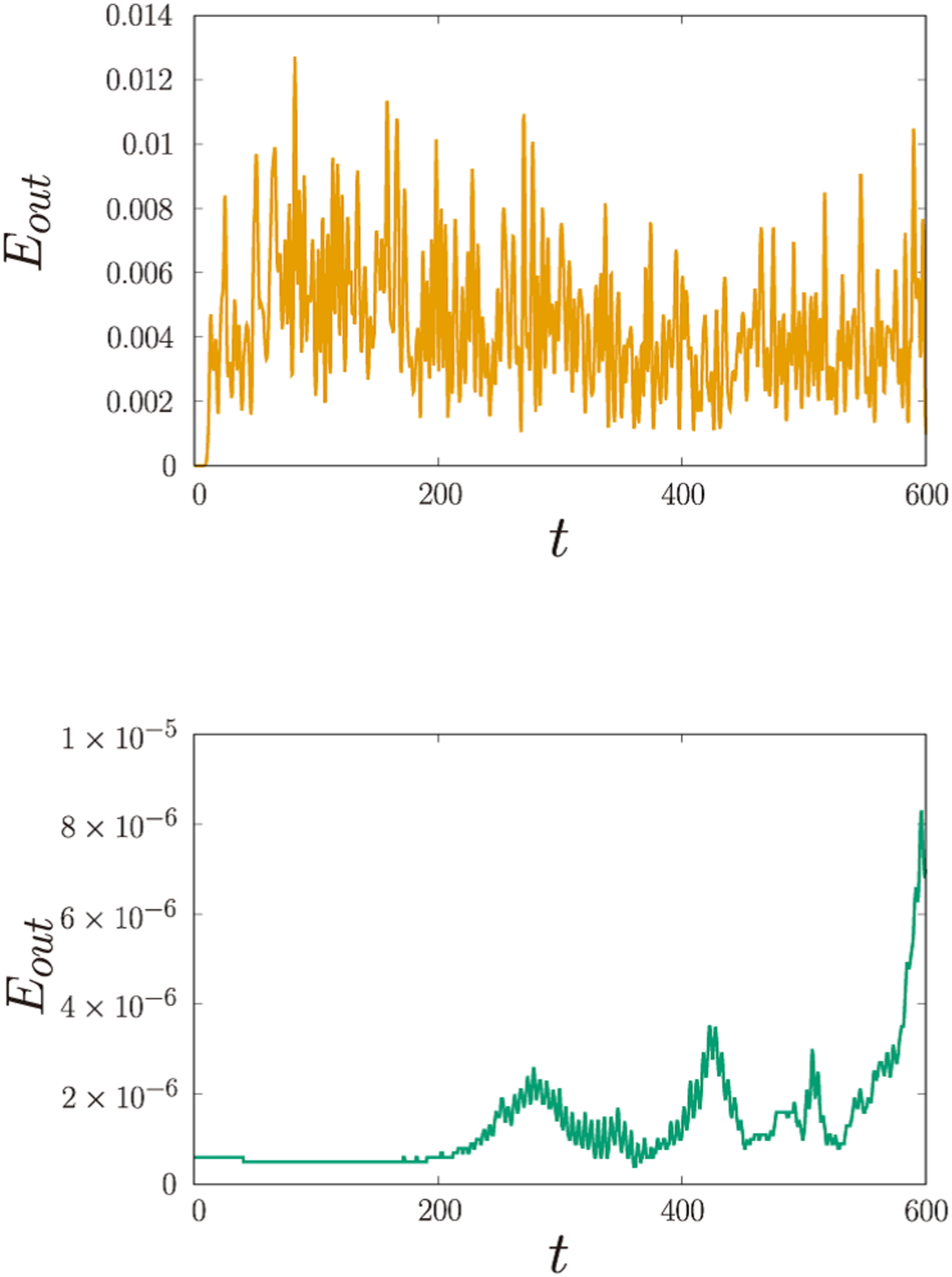} 
\end{center}
\vspace{-0.5cm}
\caption{$E_{\rm out}$ for the Higgs (upper panel) 
and confinement (lower panel) phases, respectively.
}
\label{Eout2}
\end{figure}

In the 2D case, electric flux is initially put on the central region.
In the practical calculation, the initial configuration is prepared as follows,
\be
&&\rho_{x,y}-\rho_0=-(-)^x\Delta, \;\; \mbox{for $6\leq x\leq 15$ and $y=10$}, 
\nonumber \\
&&\rho_{x,y}=\rho_0,  \;\; \mbox{otherwise}.
\label{electricF2D}
\ee
This configuration of $\{\rho_{x,y}\}$ describes the zigzag electric flux (on the gauge
lattice) extended in the $x$-direction with 10 lattice spacing.
[For more detailed dual lattice structure and definition of electric field, 
see Fig.~\ref{2Dlattice} in Sec.~\ref{disorder}.]
As in the 1D case, we add very small but finite fluctuations in local boson
density.

In Fig.~\ref{ectric2DelF}, we show the behavior of the electric flux (a)
in the confinement region for 
$J=0.01, \mu/J=2000$ and $U/J=175,V/J=30$ (MI close to DW), and also (b)
in the Higgs region $U/J=45,V/J=5$ (SF close to MI).
For the confinement region, we put $\rho_0=7$, which is the equilibrium value
for the above parameters. 
The source electric charge at $x=6$ and $15$ are $\pm \Delta=\pm 1$, respectively.
Even for the smallest unit charge, the electric flux is stable up to 
$t=300(= 3\times \hbar/J)$,
and it gradually decays after that. 
Here, the hopping time in our model is $\sim 2\times \hbar/J$. 
Its small decay starts to occur a little beyond the hopping time.
For larger source charges such as $\Delta=3$, the electric flux is quite stable.
On the other hand for the Higgs region, we put $\rho_0=30$, which is again
the equilibrium value for the above parameters.
We show the calculations for $\Delta=3$.
Even for this relatively large value of $\Delta$, the electric flux
breaks after a very small period, and it spreads whole
region and fluctuates strongly.
In Fig.~\ref{Eout2}, we also show the square of the electric flux outside of the region
$\vec{r}\neq (6\leq x\leq 15,y=10)$, $E_{\rm out}$, which is defined
similarly to the 1D case in Eq.~(\ref{defEout}).
The results in Fig.~\ref{Eout2} obviously support the results in Fig.~\ref{ectric2DelF}.

\begin{figure}[h]
\centering
\begin{center}
\includegraphics[width=6cm]{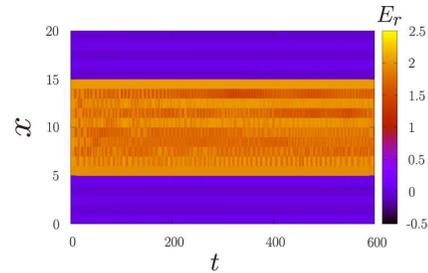} 
\end{center}
\vspace{-1cm}
\caption{Time evolution of electric flux put in the center of the system.
$J=0.05, \rho_0=7$ and $\Delta=2$.
Unit of time is $\hbar/(20J)$.
Electric flux is stable for long period.
}
\label{ectric2DelF2}
\end{figure}
\begin{figure}[h]
\centering
\begin{center} 
\includegraphics[width=9cm]{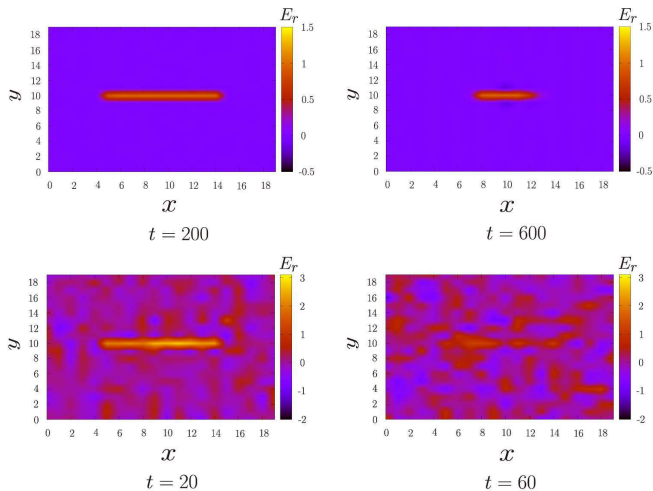} 
\end{center}
\caption{Upper panels: Time evolution of  DW-type density modulation
corresponding to electric flux put in the confinement phase
(MI close to DW).
$J=0.01, \rho_0=7$ and $\Delta=1$.
Electric flux shortens but is stable for long period.
Lower panels: Time evolution of electric flux put in the Higgs phase (SF close to MI).
Electric flux `melts' and its fluctuations start immediately.
}
\label{ectric2Dwhole}
\end{figure}

In order to verify the universality of the above result, we investigated various
parameter regions of the EBHM.
In particular, the stability of the electric flux is a very important phenomenon.
In Fig.~\ref{ectric2DelF2}, we show the calculations of the case of relatively large
hopping $J=0.05$ and $U/J=175, V/J=30, \mu/J=2000$, which corresponds to
point (a) in Fig.~\ref{2DPD}. 
The parameter point is in the confinement phase.
The electric flux is again quite stable even for larger value of $J$.

In Fig.~\ref{ectric2Dwhole}, 2D profiles of the electric flux in the whole 
$20\times 20$ region are shown.
For the confinement phase for $J=0.01$ (Fig.~\ref{ectric2DelF}), 
the electric flux starts to get shorter at $t\simeq 300(= 3\times \hbar/J)$.
For the Higgs phase in Fig.~\ref{ectric2DelF}, the electric flux decays quite rapidly.
The initial electric flux `melts' and fluctuation of electric field (i.e., particle density)
starts to develop immediately.
Time evolution of the electric field located out of the original place,
$E_{\rm out}$, in Fig.~\ref{Eout2} again supports
the above behavior.


\section{Glassy dynamics and effect of quenched disorder}\label{disorder}

In the previous section, we observed that the electric-flux string is quite stable
in the confinement phase. 
Then, it is interesting to study how higher-energy states of the DW type evolve
in that parameter region.
To examine effects of a random chemical potential on this phenomenon
is also an important problem. 
In real experiments in an optical lattice, a similar random chemical potential can 
be  implemented by using a laser speckle \cite{Schulte,Clement}.
Closely related phenomenon to the above was recently investigated by experiments 
on ultra-cold atomic gases, and it was observed that life time of states 
with higher energies is lengthened by the quenched disorder
induced by the random chemical potential~\cite{exper1}.
To reveal the origin of this glassy phenomenon and its relation to
the MBL is an interesting problem.
For the $(1+1)$D quantum electro-dynamics (QED), in which electron is 
always confined, an extremely slow evolution of entropy was observed
for configurations with background charges~\cite{2DQED}.
In this section, we focus on the 2D model and study if the confinement phase 
exhibits glassy dynamics,
and if so, we clarify its origin from the gauge-field point of view.

Parameters of the numerical studies in this section are $J=0.05, U/J=175, V/J=30$
(confinement region) and the average particle density $\rho_0=7$.
We employ the random chemical potential distributed uniformly as 
$\mu_i\in \mu\pm[-{W\over 2}, +{W\over 2}]$ with $\mu=100$, and
study the model with some specific values of $W$.
Unit of time is $\hbar/(20J)$ in this section as we put $J=0.05$ in the practical
calculation.

\begin{figure}[t]
\centering
\includegraphics[width=7cm]{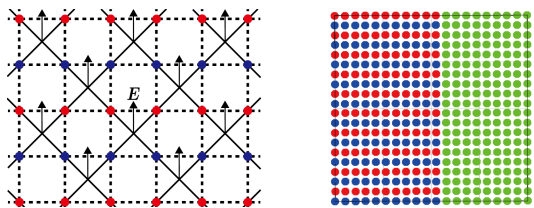} 
\caption{Left panel: The dotted lines indicate the original square optical lattice,
and the solid lines the gauge lattice.
Upward arrows indicate electric flux $\mathbf{E}$.
Right panel: Initial density configuration is shown.
Particle numbers are $9$ (blue), $7$ (green) and $5$ (red), respectively.
The left half is filled with a finite synthetic electric field as shown by the left panel.
}
\label{2Dlattice}
\end{figure}

\begin{figure}[h]
\centering 
\includegraphics[width=7.5cm]{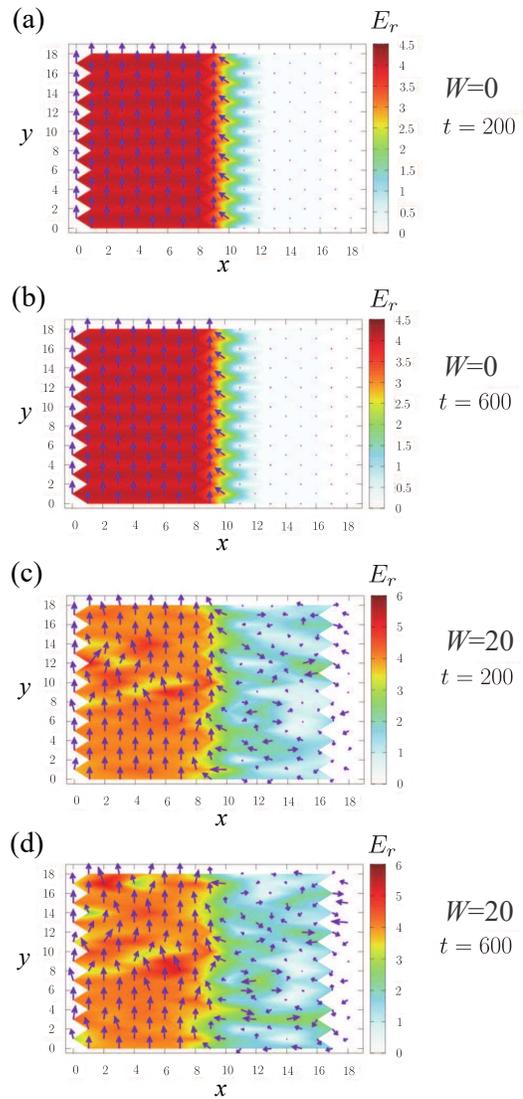} 
\caption{(a) and (b):
Time evolution of electric field $\mathbf{E}$ in the confinement phase without disorder.
Direction of arrows indicate the electric field direction, and length of arrows
and color show its magnitude.
Electric field is quite stable and keeps its original configuration.
(c) and (d): Confinement phase with disorder $W=20$.
Electric filed spreads from the original region.
There, obtained are data for a single initial configuration.
}
\label{ectric2Dhalf}
\end{figure}

First, we study time evolution of the initial state in which 
a half of the system is a DW-type configuration,
and the other half is a homogeneous state with $\rho_i\simeq \rho_0=7$.
More precisely, the DW-type region of the initial configuration is the state filled with 
electric field pointing to the $y$-direction.
Detailed lattice structure of the gauge system, electric field defined on links of 
the gauge lattice, and the initial configuration are shown in Fig.~\ref{2Dlattice}.
Time evolution of this kind of configurations is a good measure for the stability
of a bunch of electric flux tubes.

In Fig.~\ref{ectric2Dhalf}, we show the time evolution of the 2D system 
with $W=0$ and $W=20$.
Data are obtained for a single initial-configuration realization.
Other samples give almost the same results.
It is obvious that in the case of $W=0$, the electric field is quite stable. 
This is an expected result from the stability of electric flux in the confinement phase.
On the other hand, in the case of $W=20$, it tends to spread out the empty space.
We examined the case with $W=10$ and $W=30$, and obtained similar results.
That is, the EBHM and GHM exhibit glassy dynamics in the case without disorder,
and disorder hinders glassy nature.
This is somehow an `unexpected' result.
Disorder often enhances the localization even if there are interactions between
particles.
However as we explain later, the gauge-theoretical view point gives a clear interpretation 
to above phenomenon.
Before going into discussion on this point, let us consider another example of
glassy dynamics of the present system.

\begin{figure}[h]
\centering
\begin{center} 
\includegraphics[width=6cm]{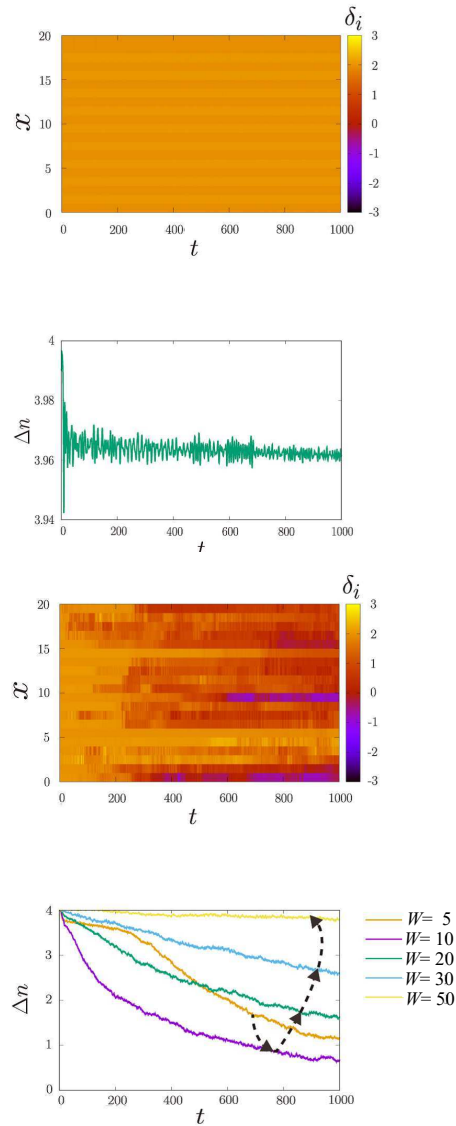} 
\end{center}
\vspace{-0.5cm}
\caption{Upper panel: Time evolution of the local DW order $\{\delta_i\}$ at $y=10$
in the confinement phase with $W=0$.
DW order is stable for long period.
Second panel: The density difference keeps $\Delta n\simeq 4$.
Third panel: Time evolution of $\{\delta_i\}$ at $y=10$ for $W=20$. 
$\{\delta_i\}$ starts to fluctuate after certain period.
Bottom panel: $\Delta n$ as a function of time for various $W$s.
The result indicates the existence of critical $W$ at which the DW 
fades away maximally.
The dotted arrow indicates an increase of $W$ as guide of eyes.
There, obtained are data for a single initial configuration.
}
\label{randomwhole}
\end{figure}

Next, we consider the evolution of the initial configurations of the genuine 
DW type such as $\rho_i=\rho_0+(-)^i\delta$ in the whole system
[$(-)^i=1 (-1)$ for even sites (odd sites)].
From the gauge-theoretical point of view, this configuration is nothing but
the state filled with a bunch of electric flux tubes pointing to opposite 
directions alternatively.
From the above observation indicating the stability of the uniform electric field 
in the confinement phase, how the genuine DW state evolves is an interesting
problem, and if this configuration is stable, we can conclude that the confinement
phase has the genuine glassy dynamics.

For $\delta=2$, calculations of the local DW order parameter
$\delta_i\equiv (-)^i(\langle\rho_i\rangle-\rho_0)$ and the difference between
the average particle numbers at even-odd sites, $\Delta n$ in Eq.~(\ref{OPs}), 
are shown in Fig.~\ref{randomwhole} for various $W$s.
There, obtained are data for a single initial configuration. 
Other samples give almost the same results.
The calculations show an interesting phenomenon.
For $W=0$, i.e., the case without disorders, the density difference $\Delta n$
keeps the original value $\Delta n\simeq 4$ for a long period.
On the other hand, 
snapshots of $\{\delta_i\}$ in the central region of the system shown in
Fig.~\ref{randomwhole} indicate that the system of $W=20$ evolves towards 
the state of $(\bar{\rho}_e-\bar{\rho}_o)\simeq 0$ with local density fluctuations,
whereas the system $W=0$ is quite stable.
The calculation of $\Delta n$ in Fig.~\ref{randomwhole} shows that
as $W$ increases from $0$ to $10$, the stability of the initial state decreases.
However, further increase of $W$ makes the DW state stable.
Although analysis for larger times is needed to conclude
$\Delta n\neq 0$ for $t\to \infty$,
this behavior of the disorder system is reminiscent of the MBL dynamics. 
Furthermore, it is expected that there exists a critical value of
$W$, $W_c\sim 10$, at which the DW fades away maximally~\cite{Luitz,Pal}.
This interesting observation will be discussed in Sec.~\ref{discussion}
from the gauge-theoretical point of view. 
Here, we report that we recently observed somewhat similar phenomenon
in quasi-1D system, Creutz ladder model, by the exact diagonalization~\cite{ours6}.
Without the NN repulsion and disorder, the system exhibits flat-band localization.
In the presence of the NN interaction, localized states survive.
As increasing disorder in chemical potentials, the localization properties of the system
weaken at intermediate disorder regime.
Further increase of disorder makes the system localized again, i.e.,
a DW-type modulation fades away maximally at intermediate disorder strength.
We discussed there that a crossover from the flat-band Anderson localization to MBL
takes place under increase of disorder.


\section{Discussion and conclusion}\label{discussion}

In this paper, we studied the 1D and 2D EBHM by the GW methods
from the view point of the quantum simulation for the gauge theory.
We first clarified the phase diagrams at relatively large fillings 
$\rho_0=(7\sim 30)$.
The phase diagrams themselves exhibit a rather interesting structure 
composed of the SF, MI, DW and SS phases.
We identified the parameter regions in the phase diagrams corresponding to  
the confinement and Higgs phases.
Then, we studied the time evolution of configurations with electric flux tube,
and verified that electric flux tube is stable in the confinement phase but 
breaks immediately in the Higgs phase.
The stability in the confinement phase increases as the magnitude of charge
at the edges of the electric flux, $\Delta$, increases.  

After the above observations, we studied the effect of disorder caused by 
the random chemical potential 
in the 2D system, which generates density inhomogeneity in the system.
We first verified the stability of the electric field in the confinement phase by
studying time evolution of electric field filling half of the system.
For the case without disorder, the electric field remains stable for long periods.
Then, we introduced disorder and found that disorder induced by the random chemical
potential renders the electric filed unstable.
This is somehow an `unexpected result', but is plausible from the gauge-theoretical
point of view.
In the gauge theory, it is established that the quark confinement takes place
as a straight electric flux tube forms between a quark-anti-quark pair and 
it exhibits almost no fluctuations.
This confinement picture explains the stability of the electric field filling
the half of the system in the case of without disorder.
On the other hand, 
the random chemical potential induces spatial electric field fluctuations as
it generates inhomogeneous background charges. 
The above picture obviously is based on the Gauss law, $\nabla\cdot \vec{E}=Q_e$,
where $Q_e$ is charge density.
In the quantum simulations of the GHM, the NN repulsion, as well as the exact 
gauge symmetry, plays an essential role for the Gauss-law constraint to be satisfied.
Recently, some related observation with the above was given in Ref.~\cite{string},
in which the experiments on Rydberg atom chain in Ref.~\cite{Rydberg} was
interpreted in the gauge-theory framework.
There, the strong NN repulsion of the Rydberg state hinders its occupation
on NN sites, and this constraint of the Hilbert space can be regarded as the
Gauss law.
The emergent gauge invariance explains the very slow dynamics observed in
Ref.~\cite{Rydberg}.

In other words, the glassy dynamics in the confinement phase observed in the present
work results from strong interactions between atoms in the confinement
regime.
It was shown in Ref.~\cite{carleo} that MBL and glassy dynamics appear
due to frustrating dynamical constraints by interactions.
 
As the second case with disorder, we investigated the stability of the genuine DW 
configuration with high energies in the whole system.
Interesting enough, we found that  the DW  configuration 
is stable for a long period in the case {\em without disorder}, whereas
the inhomogeneous particle density caused by {\em moderate} $W$s reduces 
the robustness of the DW-type configuration.
This result means that the disorder-induced spatial density modulations 
hinder the glassy dynamics.
Obviously, this behavior is reminiscent of the quark confinement mechanism
explained above.

Interestingly enough, we observed that further increase of disorder makes 
the DW-type configurations tend to dynamically survive again.
Recently, certain related experiment was performed on the ultra-cold
atoms and similar result was obtained, i.e., the random chemical potential
enhances life time of the high-energy configurations~\cite{exper1,numerical}.
This may be an expected result, i.e., disorder enhances the localization. 
For stronger disorders with $W>W_c$ in the present system, background charge is
modulated strongly, and as a result, the gauge-theoretical picture does not 
work anymore.  
We think that the EBHM in the present parameter regime exhibits interesting multiple
`phase transitions' or `crossovers' from the interaction induced glassy dynamics to 
the ordinary MBL by increasing the strength of disorder and in between regime
an ergodic phase exists.

From the above observations, it is interesting to study atomic gas systems
with NN repulsions by varying strength of disorder.
We expect that similar experiment on them to those in Ref.~\cite{exper1} sheds light on
our picture of the glassy dynamics of the confined gauge theory obtained
in this work.
It is also important to examine effects of disorders in the experiments
on the Rydberg atom chain~\cite{Rydberg} by introducing disorder in detuning
and/or Rabi frequency.
We expect that similar `phase transitions' are to be observed there.

Also, a recent numerical study \cite{Sierant} suggested that the glassy dynamics 
(slow down to the relaxation) is related to MBL. 
In our work, as shown in Fig.~\ref{randomwhole} (d), the tendency of the glassy 
dynamics becomes stronger as weaker disorder in the confinement phase. 
That is, we expect that the confinement phase has also the MBL properties. 
In confinement phase, such a conjecture has been verified for other lattice gauge
models \cite{2DQED,Smith1,Smith2}.

As related subject, we would like to comment on works in which localization of
magnetic flux lines was studied for the type-II superconductors~\cite{nandkishore,pretko}.
Although type-II superconductors correspond to the Higgs phase of the gauge theory,
there exists duality between confinement and superconductivity.
Confinement of the gauge theory takes place as a result of condensation of 
magnetic charges such as a magnetic monopole.
Squeeze of electric flux in the confinement phase is sometimes called dual 
Meissner effect~\cite{Man}.
Therefore, the localization of magnetic flux lines in superconductors suggests
the similar localization of electric flux string in the confinement phase, which 
is observed in this work.

Here, let us comment on reliability of the time-dependent GW methods.
We summarize the technical aspects first.
We study time evolutions up to $t\sim 1000$.
From unit of time, this corresponds to $t\sim 10\hbar/J \ 
(\mbox{or } 50\hbar/J)$.
Furthermore, as we used the forth-order Runge-Kutta formula, generated
errors are $O(dt^5)$ with time slice $dt=10^{-5}$.
Contrary to the time-dependent DMRG and TEBD, truncations of quantum states during
evolutions are not done.
Unfortunately, no reliable numerical methods exist for examining correctness of 
our calculations, in particular for 2D systems and even for 1D systems of high fillings.
However, experiments on ultra-cold atoms, i.e., quantum simulations,
provide useful information on the reliability of the time-dependent GW methods.

As far as we know, there are at least two experiments, which are useful for
the present examination.
First one is the experiment on ultra-cold Bose gases in a 2D disordered optical
lattice~\cite{exper1}, which we have already mentioned in the main text.
There, time evolution of Bose gas filled in a half of the optical lattice was observed
in the presence of on-site disorder in order to study glassy dynamics and MBL.
Corresponding to the above experiment, numerical simulations by the
time-dependent GW methods were performed for time evolutions in rather long
times~\cite{numerical}. 
Obtained results are in good agreement with the experiments.
Second one is study on quench dynamics of ultra-cold atoms in a 2D optical lattice.
In order to examine Kibble-Zurek scaling for a quantum phase transition,
quench dynamics from Mott to SF was observed by experiments~\cite{exper2}. 
In particular, scaling exponents were estimated from experimental data.
Stimulated by this experiment, we simulated the above quench dynamics 
by the time-dependent GW methods~\cite{ours5}.
We found that the experiments and our numerical simulations are in good
agreement. 
We also showed that the measurements in the experiment were performed
in rather late times far apart from the phase transition time, and it is source of 
the discrepancy with the Kibble-Zurek scaling exponents and the observed ones.
Even though above two are both studies on low-filling systems, they gave positive
evidences for reliability and applicability of the time-dependent GW methods to 
time evolution for long times.
As we studied fairly large-filling systems in this work, we think that there are
sufficient grounds to believe that the obtained results by the time-dependent
GW methods are correct.

In the present work, we employed the GW variational methods, we could not calculate
the entanglement entropy, which is often used for identification of MBL.
In the previous paper~\cite{ours7}, we discussed reliability of the GW methods
comparing it with quantum Monte-Carlo simulations.
As emphasized there, quantum correlations are taken into account by 
the GW methods in some amounts.
Unfortunately, their precise estimation is not known yet, and investigation by
means of concrete models are obviously welcome.
One example is a bosonic version of the Creutz ladder model, 
which can be viewed as a gauge-Higgs model on the ladder.
Relatively large-filling cases can be studied by both the truncated Wigner method and
GW methods.
We expect that the glassy dynamics takes place there as in the original
fermionic Creutz ladder model~\cite{ours6}, and the glassy dynamics
of the confined gauge theory is closely related to MBL.
We are now planning a study on it, and we hope that the results will be reported 
in the near future.



\begin{thebibliography}{99}
\bibitem{FN1}
More precisely, the next-NN (NNN) repulsions have to be added to $H_{\rm EBH}$
in Eq.~(\ref{HEBH}) for the quantum simulation of the genuine GHM in 2D.
However, low-energy system derived from  $H_{\rm EBH}$ in Eq.~(\ref{HEBH}) 
is gauge-invariant, and the universality class does not change by the existence of 
the NNN repulsions.
\bibitem{QMC}
Phase diagram of the 1D EBHM for a large filling case was obtained by the path-integral
quantum Monte-Carlo simulations (QMCS) in our previous work.
[Fig.~12 in Ref.~\cite{ours4}.]
The QMCS phase diagram is in good agreement with Fig.~\ref{1DPD},
while the locations of the phase boundaries cannot be compared directly
as $\Delta \tau$ is set unity in the path-integral QMCS.
More precise QMCS is possible only for low fillings.
See, for example Ref.~\cite{Batrouni}.
\end{thebibliography}
\end{document}